\pdfoutput=1

\documentclass[11pt]{article}

\usepackage{acl}

\usepackage{times}
\usepackage{latexsym}
\usepackage{hyperref}
\usepackage{dblfloatfix}
\usepackage[T1]{fontenc}

\usepackage[utf8]{inputenc}

\usepackage{microtype}

\usepackage{graphicx}
\usepackage[notransparent]{svg}
\svgpath{{img/}}

\title{Detection of tortured phrases in scientific literature}

\author{Eléna Martel, Martin Lentschat, Cyril Labbé \\
    Univ. Grenoble Alpes, CNRS, Grenoble INP, LIG, 38000 Grenoble, France \\
    \texttt{\{martin.lentschat,cyril.labbe\}@univ-grenoble-alpes.fr}
    }

\begin{document}
\maketitle
\begin{abstract}
This paper presents various automatic detection methods to extract so called {\it tortured phrases} from scientific papers.
These tortured phrases, e.g. {\it flag to clamor} instead of {\it signal to noise}, are the results of paraphrasing tools used to escape plagiarism detection.
We built a dataset and evaluated several strategies to flag previously undocumented tortured phrases.
The proposed and tested methods are based on language models and either on embeddings similarities or on predictions of masked token.
We found that an approach using token prediction and that propagates the scores to the chunk level gives the best results.
With a recall value of $.87$ and a precision value of $.61$, it could retrieve new tortured phrases to be submitted to domain experts for validation.
\end{abstract}

\section{Introduction}

Over the past few years, the research community has been confronted with an emerging issue related to the use of content rewriting tools. 
These tools are being used to hide crude plagiarism.
Some of these rewriting tools, called {\it spinners}\footnote{SpinBot (\url{https://spinbot.com}), SpinnerChief (\url{https://www.spinnerchief.com})}, used to destroy the meaning of the rewritten text.
In their pursuit of publication and the relentless pressure to '\textit{publish or perish}', some researchers turn to these tools. However, these spinners, leave behind lexical traces as they transform text, replacing words with synonyms that may be less appropriate during the modification process.

For scientific text, the most brutal modifications were concerning poly-lexical sequences that carry a specific meaning as well-established scientific expressions: e.g. {\it Artificial intelligence, big data} or {\it Randomized control trial}.
By performing a 'word by synonyms' replacement, the first generation of spinners would destroy the meaning conveyed by these typical collocations. For example, the previously mentioned expressions could be tortured into {\it man-made consciousness, enormous information} or {\it randomized controlled preliminary}. 
We define a tortured phrase as \textbf{an expression resulting from the use of a spinner on a well-established scientific expression with a specific and fixed meaning}.
Its counterpart is here called expected phrase (i.e. the original scientific expression).

\citet{callfor} reveals that such meaningless expressions, referred to as {\it tortured phrases}, can actually be found in many scientific papers.
These tortured phrases not only constitute evidences of the lack of reliability and relevance of these papers, but can also be used to quickly retrieve articles that are thus suspected of having employed spinners.
A manually collected set of tortured phrases is used as {\it fingerprints} \citep{CabanacLabbe2021} by the {\it Problematic Paper Screener}\footnote{\url{https://www.irit.fr/~Guillaume.Cabanac/problematic-paper-screener}} to comb the scientific literature for such problematic papers. 
The authors are querying the academic search engine \href{https://www.dimensions.ai/}{Dimensions.ai} \citep{herzog2020dimensions} to retrieve articles with known tortured phrases.

The set of manually collected tortured phrases is limited to the expertise of its contributors. 
Tortured phrases from many scientific fields are still to be listed as fingerprints, so to be able to flag undetected problematic papers. 
To this date (13 oct. 2023), 11.945 papers containing tortured phrases have been flagged by the website {\it Problematic Paper Screener}, with more to come as the number of known tortured phrases increases.
While it's possible that with the context of 2023, Large Language Models can perform paraphrasing of higher quality than spinners, it's crucial to note that these papers have already been published and remain accessible.
Also, amongst the 12k flagged articles, 1278 have been published in 2023, as well as 2 articles to be published in 2024.
Therefore, this problem still remains and it is of paramount importance to identify them for retractions.

This paper aims at testing automatic methods to distinguish differences between tortured phrases and expected ones. 
The main aims being to automatically identify tortured phrases that are yet not listed.
For this purpose:
\begin{itemize}
    \item We built a data set aiming at testing detection methods.
    \item We report results achieved when using different techniques that do not require massive use of labeled data, as such a large data set does not exists yet. 
    \item We explore the use of large language model embeddings, similarity measures, masking and prediction methods to flag automatically tortured phrases not previously known.
\end{itemize}

The remainder of this paper is organized as follows:
Section~\ref{sec:ReWo} discuss related work around spinners and the detection of tortured phrases.
Section~\ref{sec:data} describes the way we built our new data set.
Section~\ref{sec:expe} presents various methods and experiments for which Section~\ref{sec:results} provides detailed results.
Finally, Section~\ref{sec:conc} concludes and gives some perspectives on the task at hands.



\section{Related Work}
\label{sec:ReWo}


Spinners are capable to create several versions of an original text by substituting synonyms and altering sentence structure \citep{shahid2017accurate}.
An example can be taken from the following sentence: \textit{'The cat is eating its food.'}, which could be transformed into: \textit{'The feline is savoring its meal.'}.

It has been shown that content rewriting tools leave behind a trail of lexical artifacts \citep{shahid2017accurate}.
Some of these artifacts can manifest as tortured phrases, wherein the same tortured phrase might recur multiple times in place of an expected one.
Furthermore, \citet{zhang2014dspin} highlights that approximately 94\% of the vocabulary used by these tools is not regularly changed, which could explain why the same tortured phrases may reappear multiple times and thus reinforce the need for an effective detection method.

Some authors have set out with the objective of detecting spun text based on dictionaries of rewriting tools.
For instance, \citet{zhang2014dspin} relies on tokens and phrases that remain unchanged during the content rewriting process to assess the similarity between two articles, by focusing on elements that are not found in the dictionary and therefore have not been substituted.

On the other hand, \citet{wahle2022identifying} attempts to identify machine-generated paraphrased plagiarism.
They created a dataset of paraphrased content using commercial tools like {\it SpinBot} and {\it Spinnerchief}.
This dataset will encompass paraphrased texts from arXiv, student theses, and Wikipedia articles.
They employed three types of machine learning classifiers: logistic regression, support vector machines, and naive Bayes classification. 
Their task is a binary classification to mark the text as being spun or not.

We will be using the dataset of \citet{wahle2022identifying} in our study.
Given its method of fabrication, it contains many undocumented tortured phrases and is thus very valuable.
Nevertheless, to be usable for the evaluation of new tortured phrases detection methods, re-annotation at the token level is needed.
We did perform this on a small part of the dataset.

In \citet{callfor}, the authors collected data consisting of tortured expressions and their expected equivalents.
This will serve as a database of known tortured phrases with their counterparts.

The usage of embeddings to detect tortured phrases was previously explored by \citet{DBLP:conf/coling/LayLL22}.
They conclude that fixed embeddings (e.g. GloVe  \citep{pennington2014glove}), performs better than contextual ones (e.g. BERT \citep{BERT}) when using cosine similarity measure to distinguish tortured and expected phrases.
Our work goes beyond \citep{DBLP:conf/coling/LayLL22} as they only considered tortured phrases in bigrams.
We extended this method by evaluating two additional metrics, namely Manhattan distance and Euclidean distance, while also considering trigrams, which constitute a significant part within our dataset.
We also explored the usage of predictions of masked tokens to detect tortured phrases, which gave more satisfying results.


\section{Dataset}
\label{sec:data}

\citet{callfor} collected around 3,000 distinct tortured phrases thanks to the contribution of researchers and domain experts.
Then, we take advantage of the dataset provided by \citet{wahle2022identifying}, which comprises roughly 200,000 paragraphs in both their original and paraphrased forms using spiners. 
We automatically extracted, from the \citet{wahle2022identifying} dataset, sentences containing known tortured phrases.
This results in around 2,000 sentences containing known tortured phrases and approximately 4,000 sentences with their expected phrases. 
However, it is worth noting that some of the extracted sentences may potentially contain previously {\it unknown/unlisted} tortured phrases from various scientific fields, for some unfamiliar to us.
This presumption stems from the fact that these sentences have not undergone prior analysis by domain-specific researchers.
Thanks to the contributions of other researchers, we are able to flag occurrences of known tortured phrases and their expected phrases.
To ensure that our approach is not biased by the presence of unknown tortured phrases, 100 sentences were annotated using diverse sources (i.e. glossaries, scientific papers, and specialized databases).
In doing so, we aimed to determine whether scientifically established expressions not present in our dataset of expected phrases would surface, and subsequently, we verified if theses expressions had been altered during the paraphrasing process.

\section{Methodology}\label{sec:expe}

\begin{table*}[ht]
\centering
\begin{tabular}{ll|rr}
\hline
\hline
\textbf{Measures} & \textbf{Aggregation functions} & \textbf{Tortured phrases} & \textbf{Expected phrases}\\ \hline
\hline
Cosine similarity & Arithmetic mean & 0.136 (± 0.157) & 0.289 (± 0.201) \\
& Harmonic mean & 0.134 (\textit{± 1.856}) & 0.284 (\textit{± 0.581})\\
& Minimum & 0.088 (± 0.153) & 0.254 (± 0.205) \\\hline
Manhattan distance & Arithmetic mean & 42.901 (± 21.922) & 41.100 (± 20.233)\\
& Harmonic mean & 42.714 (± 21.852) & 40.936 (± 20.171)\\
& Minimum & 40.898 (± 21.408) & 39.427 (± 19.759) \\\hline
Euclidean distance & Arithmetic mean & 5.416 (± 2.765) & 5.184 (± 2.554)\\
& Harmonic mean & 5.391 (± 2.756) & 5.16 (± 2.546) \\
& Minimum & 5.159 (± 2.700) & 4.973 (± 2.494)\\
\hline
\hline
\end{tabular}
\caption{Average similarity and distance measures depending on the aggregation function}
\label{tab:measures}
\end{table*}

In this section, we present our methodology for the experiments involving word embedding similarity measures and the prediction of masked tokens to compare tortured phrases and expected phrases.

For the word embedding approach, cosine similarity and distance metrics were computed between the tokens of tortured phrases and the tokens of expected phrases.
The two values were then compared.
The aim of using the word embedding was to determine whether similarity and distance metrics could effectively distinguish the two classes of phrases.
The underlying idea is that expected phrases, being conventional and legitimate, would obtain higher similarity scores and lower distance metrics scores, reflecting greater semantic coherence and regularity compared to tortured phrases.


Bigrams and trigrams were compared by, first calculating scores between constituent bigrams, then aggregating the two scores via arithmetic mean, harmonic mean or minimal value.
For example, for the bigram \textit{'big data'}, the three measures were applied between the two tokens.
For a trigram like \textit{'support vector machine'}, the measures were computed between all bigrams pairs : \textit{'support'} \& \textit{'vector'}, \textit{'support'} \& \textit{'machine'}, \textit{'vector'} \& \textit{'machine'}.
The resulting scores were then aggregated.
Minimum takes the lowest score, mean calculates the average, and harmonic mean weights lower scores more strongly.

The chosen word embeddings are the ones from the GloVe model \citep{pennington2014glove}.
Specifically, we utilized the pre-trained 'glove-wiki-gigaword-100' model, which had shown good performance in previous work \citep{DBLP:conf/coling/LayLL22}.
For these experiments, we used the dataset containing around, 2763 tortured phrases and expected counterparts.
The dataset is out-of-context, meaning the phrases are extracted from their original sentences.
If a token within a phrase is not present in the vocabulary, no calculation is performed.

Since the semantic of a tortured phrase is destroyed during spinning (i.e. compared to the semantic of a expected phrase), we though of using language models to try to predict tokens in the text.
For this masking approach, the SciBERT~\citep{beltagy2019scibert} pretrained language model was used to predict masked words based on surrounding context.
The masking approach was inspired by the methodology used in \citet{gehrmann2019gltr}.
Specifically, we adopted their use of three metrics: probability of the original word, rank of the original word in the predicted distribution and entropy over the predicted token distribution.
Our goal was to analyze whether there were significant differences in probability, ranking, and entropy between expected and tortured phrases

Two evaluations were performed, token-level and noun chunk-level, to thoroughly analyze approach performance on detecting tortured phrases.
Tokens were labeled as 0 or 1 for classification. O when the token is not part of a tortured phrases and 1 when the token is part of a tortured phrase. 
An optimal threshold was determined to best separate the two classes based on the predicted scores.
For the token-level evaluation, we compared the true and predicted categories matched for each token.

In contrast, when using noun chunk for classification, the approach propagates the detection of a tortured token to its chunk.
The intuition being that a noun chunk containing one tortured token can be considered in full as a tortured phrase.
\newpage
In details, results were analyzed at the noun chunk level using the following rules:
\begin{itemize}
    \item A true positive (TP) is a TP if at least one token of the chunk is labeled as tortured in both the true and predicted categories.
    \item A false positive (FP) is a FP if no tokens are tortured, but at least one is predicted as tortured.
    \item A true negative (TN) is a TN if no tokens are labeled as tortured in the chunk in either true or predicted categories.
    \item A false negative (FN) is a FN if at least one token is tortured in the chunk, but no token in the chunk is predicted as tortured.
\end{itemize}
This accounts for phrases as a single unit rather than independent tokens. 
Case examples can be found in Appendix \ref{sec:appendix}, Table \ref{tab:newtor}.

\begin{table*}[h!]
\centering
\begin{tabular}{llll||llll}
\hline
\hline
\multicolumn{4}{c||}{\textbf{With punctuation}} & \multicolumn{4}{c}{\textbf{Without punctuation}} \\
\hline
\hline
Class & Precision & Recall & F1 score & Class & Precision & Recall & F1 score\\\hline
\multicolumn{4}{c||}{\textbf{Probability}} & \multicolumn{4}{c}{\textbf{Probability}} \\
0 & 0.96 & 0.70 & 0.81 & 0 & 0.98 & 0.73 & 0.83\\
1 & 0.32 & 0.81 & 0.46 & 1 & 0.22 & 0.81 & 0.35\\ \hline
\multicolumn{4}{c||}{\textbf{Entropy}} & \multicolumn{4}{c}{\textbf{Entropy}} \\
0 & 0.93 & 0.63 & 0.75 & 0 & 0.96 & 0.64 & 0.77\\
1 & 0.25 & 0.72 & 0.37 & 1 & 0.16 & 0.73 & 0.26\\\hline
\multicolumn{4}{c||}{\textbf{Rank}} & \multicolumn{4}{c}{\textbf{Rank}} \\
0 & 0.96 & 0.73 & 0.83 & 0 & 0.98 & 0.74 & 0.84\\
1 & 0.34 & 0.80 & 0.48 & 1 & 0.23 & 0.81 & 0.36\\
\hline
\hline
\end{tabular}
\caption{Results summary of token classification with and without punctuation.}
\label{tab:ProbRankEnt}
\end{table*}

\section{Results}\label{sec:results}

Here, we present the results of our experiments.

The word embedding experiments analyzed similarity and distance metrics on bigrams and trigrams to compare tortured and expected phrases.
The hypothesis was that conventional phrases exhibit greater semantic regularity in their vector representations.
The outcomes are depicted in Table \ref{tab:measures}, which showcases the cosine similarity and distance results for the various aggregations.

While Manhattan and Euclidean distances are generally greater for tortured phrases than for expected phrases, the gaps are marginal compared to cosine similarity.
It exhibited the clearest differentiation between tortured and expected phrases based on word embeddings (cf. Appendix \ref{sec:appendix}, Figure \ref{gaus}).
Additionally, harmonic mean revealed to be a poor aggregation function due to its higher variability.
However, this approach has a long computation time which reduces its usage.
In addition, while this approach shows a distinction in the overall values between tortured and expected phrases, it is not readily applicable to individual cases (i.e. standard deviation values show a clear overlap).

The masking approach leveraged language models to predict masked words in context, assessing probability, rank, and entropy differences between phrases types.
Two levels of evaluation were conducted: token-level and noun chunk-level. 
To analyze the impact of punctuation, we first generated predictions with and without punctuation marks.
We compared the results for the three metrics probability, rank and entropy.


Table~\ref{tab:ProbRankEnt} shows the precision, recall and F1 scores for the two categories with and without punctuation.
For the expected tokens (category 0), we observe high precision and recall score both with and without punctuation.
For the tortured tokens (category 1), the precision and recall scores are lower, especially without punctuation.
This suggests that the model struggles more to correctly predict the tortured tokens.
This is in part due to a class distribution imbalance in the data (i.e. the amount of legitimate tokens far exceeds the tortured tokens), which is hard to correct as this distribution is inherent to the problem at hand.
However, the scores for class 1 improve when punctuation is present. 


\begin{table}[ht]
\centering
\begin{tabular}{lll}
\hline
\hline
Precision & Recall & F1-score \\
\hline
\hline
\multicolumn{3}{c}{\textbf{Probability}} \\
0.614 & \textbf{0.873} & 0.716 \\
\hline
\multicolumn{3}{c}{\textbf{Entropy}} \\
0.589 & \textbf{0.873} & 0.706 \\
\hline
\multicolumn{3}{c}{\textbf{Rank}} \\
\textbf{0.615} & 0.867 & \textbf{0.718} \\
\hline
\hline
\end{tabular}
\caption{Results for noun chunks}
\label{tab:noun_chunk_results}
\end{table}

Table \ref{tab:noun_chunk_results} shows the precision, recall and F1 scores at the noun-chunk level.
We observed improved scores to token-level masking without noun chunks.
We obtained an interesting recall of 0.873, showing a good capability to detect new tortured phrases, but a precision of 0.615 implying that domain experts should still filter the phrases identified.

\section{Conclusion}
\label{sec:conc}

This paper presents different methods to extract {\it tortured phrases} from scientific papers.
These tortured phrases can then be used to query academics search engine in search for problematic scientific papers.
The aim is to apply this identification method of tortured phrases to increase the existing database.

The most promising method is based on large language model token predictions propagate to their noun chunks.
It achieves a good recall (~0.87) but the precision still needs to be improved (~0.61).
This means that the detection of tortured phrases still requires some sort of manual checking by domain experts.
We also noticed that distinguishing tortured phrases from their legit counterpart can be highly contextual.
Future work could try to be more context aware and explore the use of more specific language models.

\section*{Acknowledgements}
The \href{https://nanobubbles.hypotheses.org/}{NanoBubbles} project has received Synergy grant funding from the European Research Council (ERC), within the European Union’s Horizon 2020 program, grant agreement no. 951393.

\bibliographystyle{acl_natbib}
\bibliography{custom}

\appendix
\section{Example of tortured phrases}\label{sec:appendix}

\begin{figure}[htbp]
    \centering
    \begin{minipage}[b]{1\linewidth}
        \centering
        \includegraphics[width=\linewidth]{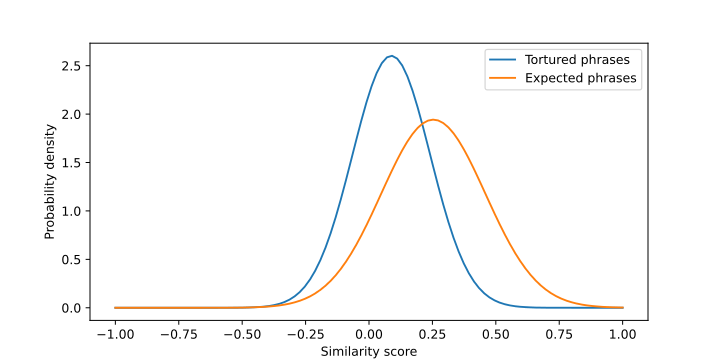}
        \caption{Cosine similarity using minimum aggregation}
        \label{gaus}
    \end{minipage}
\end{figure}

Figure \ref{gaus} shows results using cosine similarity and minimum as the aggregation function.
Table \ref{tab:newtor} shows True Positive (TP) tortured phrases detected by chunk method as well as False Positive (FP), True Negative (TN), False Negative (FN).   


\begin{table}[ht]
    \centering
    \begin{tabular}{lcl}
        \hline
        \hline
        Case &  & Decision \\
        \hline
        \hline
        & \textit{width} \textit{and} \textit{profundity} \\
        value & \hfill 1 \hspace{1em} 1 \hspace{1.5em} 1 \hspace{2em} & True \\
        predict. & \hfill 0 \hspace{1em} 0 \hspace{1.5em} 1 \hspace{2em} & Positive \\
        \hline
        & \textit{convoluted} \textit{neural} \textit{system} \\
        value & \hfill 1 \hspace{2.5em} 1 \hspace{2em} 1 \hspace{1em} & False \\
        predict. & \hfill 0 \hspace{2.5em} 0 \hspace{2em} 0 \hspace{1em} & Negative \\
        \hline
        & \textit{breast} \textit{cancer} & \\
        value & 0 \hspace{1em} 0 & True \\
        predict. & 0 \hspace{1em} 0 & Negative \\
        \hline
        & \textit{brain} \textit{tumor} & \\
        value & 0 \hspace{1em} 0 & False \\
        predict. & 1 \hspace{1em} 0 & Positive \\
        \hline
        \hline
    \end{tabular}
    \caption{Example of TP, FP, FN, TN with the chunk method}
    \label{tab:newtor}
\end{table}

\end{document}